\documentclass{article}
\usepackage{spconf,amsmath,graphicx}

\def\evec{\mathbf e}

\def\xvec{\mathbf x}

\def\muvec{\boldsymbol \mu}

\usepackage{lineno,hyperref}
\usepackage{cite}
\usepackage{amsmath,graphicx}
\usepackage{float}
\usepackage{amssymb,amsmath,comment,caption, subcaption, graphicx,amsmath,mathrsfs, array, hyperref, multirow}
\usepackage{booktabs}
\usepackage{diagbox}
\usepackage{xcolor}
\usepackage{tcolorbox}
\usepackage{todonotes}
\usepackage{lipsum}
\newtcolorbox{mybox}{fontupper=\footnotesize}

\usepackage{array}
\usepackage{amsmath}
\usepackage{mathtools}
 
\newcolumntype{d}{>{\displaystyle}c}
\modulolinenumbers[5]
\ninept

\usepackage[square,sort,comma,numbers]{natbib}
\title{CopyPaste: An Augmentation Method for Speech Emotion Recognition}
\name{Raghavendra Pappagari$^1$, Jes\'us Villalba$^{1,2}$, Piotr \.Zelasko$^{1,2}$,  Laureano Moro-Velazquez$^1$, Najim Dehak$^{1,2}$}
\address{
$^1$Center for Language and Speech Processing,
  $^2$Human Language Technology Center of Excellence, \\
    Johns Hopkins University, Baltimore, MD \\
    \{rpappag1, jvillal7, pzelasko, laureano, ndehak3\}@jhu.edu
    }

\begin{document}
\maketitle
\begin{abstract}
Data augmentation is a widely used strategy for training robust machine learning models.
It partially alleviates the problem of limited data for tasks like speech emotion recognition (SER), where collecting data is expensive and challenging.
This study proposes \textit{CopyPaste}, a perceptually motivated novel augmentation procedure for SER.
Assuming that the presence of emotions other than neutral dictates a speaker's overall perceived emotion in a recording, concatenation of an emotional (emotion $E$) and a neutral utterance can still be labeled with emotion $E$.
We hypothesize that SER performance can be improved using these concatenated utterances in model training.
To verify this, three \textit{CopyPaste} schemes are tested on two deep learning models: one trained independently and another using transfer learning from an x-vector model, a speaker recognition model.
We observed that all three \textit{CopyPaste} schemes improve SER performance on all the three datasets considered:  MSP-Podcast, Crema-D, and IEMOCAP.
Additionally, \textit{CopyPaste} performs better than noise augmentation and, using them together improves the SER performance further.
Our experiments on noisy test sets suggested that \textit{CopyPaste} is effective even in noisy test conditions.

\end{abstract}
\section{Introduction}
\label{sec:intro}

Speech emotion recognition (SER) deals with recognizing the perceived emotion of a speaker in a recording.
Emotion of a speaker can be categorized into many classes such as angry, happy, sad, neutral etc.
Recognizing the emotion automatically is challenging as it depends on many factors such as context and demographics~\cite{koolagudi2012emotion}.
Some of the applications of SER include mental health analysis~\cite{arevian2020clinical}, detecting patient's emotion ~\cite{lopez2015feature} and detecting hate speech in social media~\cite{schmidt2017survey}.

Most studies follow a two stage approach to recognize emotion from speech recordings: (1) finding a suitable speech representation and (2) building a model to predict the emotion, given the representations.
Some examples like ~\cite{huang2014speech, lim2016speech, cho2018deep, zhao2019speech}, explore the use of standard feature representations such as MFCC and OpenSMILE~\cite{eyben2013recent} features with deep learning models like CNN and LSTM. %
Few recent studies explore the possibility of learning the speech representations jointly with the emotion recognition model~\cite{tzirakis2017end, sarma2018emotion, trigeorgis2016adieu}--using raw waveform or spectrogram as the input.
While these works are focused on exploiting only datasets with emotion-related annotations or labels for SER, few other research groups explored leveraging large datasets with annotations unrelated to emotion such as phoneme and speaker identity labels~\cite{lakomkin2018reusing, pappagari2020x}.
In this respect, authors in~\cite{lakomkin2018reusing} show that knowledge learned in an ASR model trained to predict phones can be transferred to SER.
Similarly, features extracted from speaker recognition models such as x-vector model are shown to contain emotion information~\cite{pappagari2020x}.

Majority of these studies use deep learning models which thrive with more data i.e., they are data hungry~\cite{hestness2017deep}.
However, current emotion datasets are small and collecting more data with emotion labels is expensive.
Moreover, ambiguous nature of emotion makes it difficult to annotate speech recordings.
Some of the previous works attempted to tackle this problem by artificially creating more data using data augmentation techniques.
Authors in~\cite{pappagari2020x,lakomkin2018robustness} show that adding noise to the clean recordings help the model to better recognize emotions.
Altering the speaking rate of speech~\cite{lakomkin2018robustness} and vocal tract length perturbation~\cite{etienne2018cnn+} are also shown to help SER.
Few recent studies~\cite{bao2019cyclegan, rizos2020stargan} ventured into generating emotional speech features using advance techniques such as CycleGANs and StarGANs. %

In this paper, we propose a novel augmentation method for SER.
We postulate that the presence of emotions other than neutral emotion dictate the overall perceived emotion of a speaker in a recording.
In other words, if the speaker expresses an emotion other than neutral even for a short duration in a longer utterance, then that speaker is perceived to be expressing that emotion.
Consequently, concatenation of an emotional utterance with an emotion $E$ and a neutral utterance produces a new emotional utterance that can be labelled with emotion $E$.
We hypothesize that using the new concatenated utterances along with the original ones in SER model training will improve the SER performance.
We evaluate our hypothesis using two models: one model trained independently and another using transfer learning from a speaker recognition model.
We follow the framework presented in~\cite{pappagari2020x} to build the models.
We first present results with the proposed augmentation and noise augmentation, then we investigate effectiveness of these augmentation techniques in noisy conditions.

The paper is organized as follows. First, we present the proposed augmentation procedure in Section~\ref{sec:concataugment} and then the models we use to conduct experiments in Section~\ref{sec:our_models}.
We present our datasets and results in Section~\ref{sec:datasets} and~\ref{sec:results} respectively followed by conclusions in  Section~\ref{sec:conclusion}.

\section{CopyPaste}
\label{sec:concataugment}

When trying to classify the emotion of an utterance, some segments might have more emotional information than others.
To this respect, some authors observed that when an emotionally neutral speech segment and an emotional segment with emotion $E$ are played in sequence, the human listeners commonly classified the whole sequence as emotion $E$~\cite{toth2008speech}.
Therefore, the emotional segments (non-neutral) in an utterance might define listener's perception.
For example, consider a 10 s recording where the speaker is angry for the first 3 s and  manifests a neutral emotion for the remaining 7 s. 
We surmise that a human annotator might label the utterance as angry even though the speaker expresses neutral emotion for the most part of the recording. 
In these cases, the recognition of the angry emotion by machine learning models might be difficult, as the neutral emotion dominates the overall statistics of the utterance referred in the previous example.
In this work, we address this problem by proposing \textit{CopyPaste} augmentation technique.
This technique considers that a speaker is perceived to be expressing an emotion $E$ (non-neutral) even if that emotion is exhibited for a short duration.
Hence, we propose a data augmentation methodology consisting of the concatenation of an emotional utterance with emotion $E$ and a neutral utterance. The resulting concatenated utterance is then labelled with emotion $E$ for model training.
As we copy one utterance and paste (concatenate) it at the beginning or ending of another utterance to produce a new one, we have called this process \textit{CopyPaste} augmentation.
Under this method, we present three data augmentation schemes for model training:
\begin{enumerate}
    \item Concatenation of an emotional utterance (say emotion $E$) and a neutral utterance to produce another utterance with emotion $E$. We refer to this scheme as \textit{Neutral CopyPaste (N-CP)} 
    \item Concatenation of two emotional utterances with same emotion $E$ to produce another utterance with emotion $E$. We refer to this scheme as \textit{Same Emotion CopyPaste (SE-CP)}
    \item Using \textit{N-CP} and \textit{SE-CP} together during model training. We denote this scheme with \textit{N+SE-CP}
\end{enumerate}

Through \textit{CopyPaste} schemes, we can produce a greater variety in the training data which can help the model generalize better.
We expect that the \textit{N-CP} scheme i.e., concatenating emotional utterances with neutral utterances forces the model to focus more on emotional parts of an utterance.
For example, if the input is a concatenation of angry utterance and neutral utterance then \textit{N-CP} scheme forces the model to focus more on the angry part of the utterance compared to neutral part of the utterance. 
For details of \textit{CopyPaste} schemes implementation in our models and it's comparison with standard noise augmentation technique, please refer to the Section~\ref{subsubsec:copypaste_implementation} and~\ref{sec:results}.

\section{Our Models}
\label{sec:our_models}

In this section, we present the details of our SER models employed to evaluate the proposed \textit{CopyPaste} schemes. We use the same ResNet-34 architecture detailed in~\cite{pappagari2020x}, where speaker recognition models such as x-vector model are utilized for SER.

\subsection{ResNet model architecture}
In this paper, we use ResNet model reported in~\cite{villalba2019state, pappagari2020x} for utterance level representation specific to speaker.
Our ResNet model consists of three modules: frame-level representation learning network, pooling network, and utterance-level classifier.
Frame-level representation learning network operates on input frame-level features such as Mel-frequency cepstral coefficients (MFCC) and filter-bank coefficients.
We use ResNet-34~\cite{he2016deep} structure consisting of a sequence of 2D convolutional layers with residual connections between them.
The pooling network comprises a multi-head attention layer and operates on ResNet output $\xvec_t$. 
Normalized attention scores ${w_{h,t}}$ for each head $h$ are calculated as follows:
\begin{align}
  w_{h, t} = \frac{\exp(-s_h \left\|\xvec_t-\muvec_h\right\|)}{\sum_{t=1}^T \exp(-s_h \left\|\xvec_t-\muvec_h\right\|)} \;.
\end{align}

Output embedding for head $h$ is the weighted average over its inputs along the time axis:
\begin{align}
    \evec_h = \sum_t w_{h,t} \xvec_t
\end{align}
Different heads are designed to capture different speech aspects of the input signal.
We concatenate the attention heads output and pass it through a fully connected layer to obtain a single vector embedding summarizing the input.
Then, the output of the fully connected layer is passed through an utterance-level classifier to obtain model decision.
The whole network structure is illustrated in Table~\ref{tab:xvec_arch}.

\begin{table}\caption{ResNet architecture}
\label{tab:xvec_arch}
\centering
\begin{tabular}{|c|c|c|}
\hline
Component  & Layer   & Output Size   \\ \hline
\multirow{6}{*}{\begin{tabular}[c]{@{}l@{}}Frame-level\\Representation\\ Learning\end{tabular}} & $7 \times 7, 16$                                                                          & $T \times 23$        \\ \cline{2-3} 
& \begin{tabular}[c]{@{}l@{}} $\begin{bmatrix} 3 \times 3, 16 \\ 3 \times 3, 16 \end{bmatrix} \times 3$ \end{tabular}             & $T \times 23$        \\ \cline{2-3} 
& \begin{tabular}[c]{@{}l@{}} $\begin{bmatrix} 3 \times 3, 32 \\ 3 \times 3, 32 \end{bmatrix} \times 4$, stride 2\end{tabular}   & $\frac{T}{2} \times 12$    \\ \cline{2-3} 
& \begin{tabular}[c]{@{}l@{}} $\begin{bmatrix} 3 \times 3, 64 \\ 3 \times 3, 64 \end{bmatrix} \times 6$, stride 2\end{tabular}   & $\frac{T}{4} \times 6$     \\ \cline{2-3} 
& \begin{tabular}[c]{@{}l@{}} $\begin{bmatrix} 3 \times 3, 128 \\ 3 \times 3, 128 \end{bmatrix} \times 3$, stride 2\end{tabular} & $\frac{T}{8} \times 3$     \\ \cline{2-3} 
& average pool $1 \times 3$                                                            & $\frac{T}{8}$         \\ \hline
Pooling                                                                                       & 32 heads attention                                                               & $32 \times 128$      \\ \hline
\multirow{2}{*}{\begin{tabular}[c]{@{}l@{}}Utterance-level\\ Classifier\end{tabular}}         & FC                                                                               & 400           \\ \cline{2-3} 
                                                                                              & FC                                                                               & \#classes \\ \hline
\end{tabular}
\end{table}

\subsection{Speech Emotion Recognition (SER)}
\label{subsec:emotion_recog}

For achieving SER using the ResNet architecture, we train two models: one model trained with random initialization and another initialized with a model pre-trained for speaker classification.
We use 23-dim MFCC representation as input to the ResNet model.
Before passing them through ResNet, energy-based voice activity detection and mean normalization are applied.
We minimize the cross-entropy loss function during training using Adam optimizer with default parameters in PyTorch.
The epoch with the best weighted f1-score on the development set is chosen for evaluating on test set.
We report average of weighted f1-scores from 3-runs on the test set for each emotion dataset considered.

\subsubsection{Transfer learning from a model pre-trained for speaker classification}
For pre-training, we use several datasets with speaker labels.
In this work, we use VoxCeleb1, VoxCeleb2, NIST SRE4-10 and Switchboard datasets which together contain approximately 12000 speakers.
We use noise and music augmentation in pre-training to improve speaker classification performance. 
In speaker recognition community this model is commonly referred as x-vector model.
For more details, please refer to~\cite{villalba2019state}.

To transfer the knowledge learned in this model for SER, we follow the fine-tuning procedure described in~\cite{pappagari2020x} i.e., replace the final speaker-discriminative layer with one emotion-discriminative layer and fine-tune to minimize emotion loss.
In other words, we use the weights learned in pre-training for all the layers except the last layer and then optimize all the weights for emotion classification.

\subsubsection{CopyPaste schemes implementation}
\label{subsubsec:copypaste_implementation}
During training, we randomly sample a batch of 128 utterances and perform \textit{CopyPaste} based on the emotion class labels.
For \textit{SE-CP} augmentation scheme, we pick the utterances with the same emotion labels and randomly pair them for concatenation.
For \textit{N-CP} augmentation scheme, we pick utterances with neutral emotion and randomly pair them with all utterances in the batch including neutral utterances.
In this scenario, there is a risk that the resulting models are biased against the neutral emotion, as 50\% of each augmented utterance is of neutral emotion, and yet we force the model to predict the emotion of the other 50\% of the augmented utterance.
To avoid that danger, we perform \textit{CopyPaste} augmentation only for 80\% of the batches in each epoch.
With the same premises, in the \textit{N+SE-CP} scheme, we follow each of \textit{N-CP} and \textit{SE-CP} schemes for 40\% of the batches in each epoch amounting to 80\% of batches with \textit{CopyPaste} augmentation.
To avoid overfitting, we randomly pick 4 s from each recording for concatenation instead of whole recording.
We note that the average length of the training recordings in our datasets is less than 6 s.
Hence, our hypothesis is affected only with negligible likelihood by picking only 4 s of each recording for concatenation.

\subsubsection{Noise augmentation}
\label{subsubsec:noise_augmentation}
In this work, we augment the training data by adding noise and music from MUSAN corpus~\cite{snyder2015musan}.
Our augmented data contains six copies of training set with SNRs of 10 dB, 5 dB and 0 dB after adding noise and music.
We denote the models trained with clean and augmented data as \textit{Clean+Noise}.
As researchers showed that effectiveness of adding noise to the training data is more evident on noisy test data compared to clean test data~\cite{hsiao2015robust}, we compare noise augmentation with \textit{CopyPaste} in noisy test conditions.
As emotion datasets are usually clean and have higher SNR, adding noise to the test data is considered.
We create two sets of test data, one with SNR level of 10 dB and another with 0 dB for comparison with \textit{CopyPaste}.

\section{Datasets}
\label{sec:datasets}

We evaluate the \textit{CopyPaste} augmentation schemes on three contrasting datasets: MSP-Podcast (natural and no restriction on spoken content), Crema-D (acted and restricted to 12 sentences), and IEMOCAP (acted -- scripted and improvised).
The details are as follows.

\subsection{MSP-Podcast dataset}
MSP-Podcast dataset\footnote{Data provided by The University of Texas at Dallas through the Multimodal Signal Processing Lab. This material is based upon work supported by the National Science Foundation under Grants No. IIS-1453781 and CNS-1823166. Any opinions, findings, and conclusions or recommendations expressed in this material are those of the author(s) and do not necessarily reflect the views of the National Science Foundation or the University of Texas at Dallas.} is collected from podcast recordings which discuss about a variety of topics like politics, sports and movies~\cite{lotfian2017building}.
The annotations are obtained using crowd-sourced workers with a minimum of 5 workers for each utterance.
In this work, we used 5 emotion classes: angry, happy, sad, neutral, disgust for classification as in~\cite{pappagari2020x}. 
We used the standard splits in Release 1.4 for training, development, and testing.
This dataset has 610 speakers (14200 utterances) in the training split, 30 (2893 utterances) in the development, and 50 speakers (6482 utterances) in the test split.

\subsection{Crema-D Dataset}
Crema-D dataset\footnote{https://github.com/CheyneyComputerScience/CREMA-D} is a multimodal dataset (audio and visual) containing utterances with enacted emotions by 91 professional actors from a pre-defined list of 12 sentences.
This dataset includes 48 male and 48 female actors with diverse ethnicity and age distribution.
We follow the same data splits as in~\cite{pappagari2020x} with 51 actors for training, 8 for development and 32 for testing.
We experiment with 4 emotion categories: angry, happy, neutral and sad.

\subsection{IEMOCAP}
IEMOCAP dataset is a multimodal conversational dataset recorded with 5 female and 5 male actors~\cite{busso2008iemocap}.
It contains improvised and scripted conversations about pre-defined topics in 5 sessions.
Each utterance is annotated by at least 3 annotators who were asked to choose from 8 emotion categories namely angry, happy, neutral, sad, disgust, fear and excited.
In this work, we choose a subset of data with labels angry, happy, neutral and sad as in the previous works.
We perform 5-fold cross-validation (CV) on this dataset as it is relatively small and contains only 10 speakers.
For model training on each fold, we use three sessions for training the model, one session for development, and the remaining session for testing.
In each fold, we use weighted f1-score as our metric, and hence, we report an average of weighted f-scores of 5-fold CV for each experiment.

\begin{table}\caption{SER results (weighted f1-scores) with randomly initialized ResNet model. \textit{Clean+Noise} and \textit{Clean} denote SER model training is on clean and noise augmented data, and clean data respectively. In parenthesis, absolute improvement compared to model trained without \textit{CopyPaste} (No \textit{CP}) is shown.}
    \label{tab:random_init_resnet_emotion_results}
    \centering
    \resizebox{\columnwidth}{!}{
    \begin{tabular}{l|l|l|l|l|l}
      \toprule
    Dataset    &Emotion data	&		No \textit{CP}	&	\textit{SE-CP} & 	\textit{N-CP}	&	\textit{N+SE-CP} 	\\
        	\midrule
    \multirow{2}{*}{MSP-Podcast} &  Clean  	&	47.36	&	48.34	&	49.14 &	\textbf{49.69}	(+2.33)	\\
    & Clean+Noise &	48.15	&	50.61 &	49.25		&\textbf{50.71} (+2.56)	\\
    \midrule
    \multirow{2}{*}{Crema-D}  & Clean &	71.46	&	71.80	&	\textbf{74.34} (+2.88) &	73.79		\\
    
     & Clean+Noise	&	70.59	&	72.83	&	\textbf{75.87} (+5.28) &	74.55		\\
    \midrule
    \multirow{2}{*}{IEMOCAP}  & Clean 	&	43.03	&	45.84	&	44.19	&	\textbf{45.88} (+2.85)	\\
    & Clean+Noise &	43.65	&	49.49	&	\textbf{52.34} (+8.69) &	51.41		\\
    \bottomrule
    \end{tabular}
    }
\end{table}

\begin{table} \caption{SER results (weighted f1-scores) with ResNet model pre-trained for speaker classification. \textit{Clean+Noise} and \textit{Clean} denote SER model training is on clean and noise augmented data, and clean data respectively. In parenthesis, absolute improvement compared to model trained without \textit{CopyPaste} (No \textit{CP}) is shown.}
    \label{tab:pretrained_resnet_emotion_results}
    \centering
    \resizebox{\columnwidth}{!}{
    \begin{tabular}{l|l|l|l|l|l}
      \toprule
    Dataset    &Emotion data	&		No \textit{CP}	&	\textit{SE-CP} & 	\textit{N-CP}	&	\textit{N+SE-CP} 	\\
        	\midrule
    \multirow{2}{*}{MSP-Podcast} &  Clean  	&	56.79	&	\textbf{58.68} (+1.89) &	57.71 	&	58.22	\\
    & Clean+Noise 	& 57.91	&	\textbf{58.62} (+0.71)	&	57.82 &	58.13		\\
    \midrule
    \multirow{2}{*}{Crema-D}  & Clean 	& 77.86	&	78.54	&\textbf{80.18} (+2.32) &	79.21		\\	
    
     & Clean+Noise 	&	79.60	&	79.98	&	\textbf{80.17} (+0.57) &	79.88		\\
    \midrule
    \multirow{2}{*}{IEMOCAP}  & Clean 	& 61.18	&	\textbf{62.15} (+0.98)	&	61.21 &	61.90		\\
    & Clean+Noise 	&	62.57	&	63.08	&	63.48	&	\textbf{63.78} (+1.21)	\\
    \bottomrule
    \end{tabular}
    }
\end{table}

\begin{table}\caption{Class-wise f1-scores on Crema-D dataset with \textit{CopyPaste (CP)} schemes. We used ResNet model pre-trained for speaker classification and trained on clean data; No \textit{CP} denotes model trained without \textit{CopyPaste}}
    \label{tab:classwise_emotion_results}
    \centering
    \resizebox{\columnwidth}{!}{
    \begin{tabular}{l|l|l|l|l}
      \toprule
        Emotion class   	&		No \textit{CP}	&	\textit{SE-CP} & 	\textit{N-CP}	&	\textit{N+SE-CP} 	\\
        	\midrule
    \textit{Sad}     &	20.48	&	20.59	&	21.11	&	\textbf{22.37}	\\
    \textit{Happy}   &	37.17	&	46.19	&	\textbf{54.47}	&	46.7	\\
    \textit{Angry}   &	70.53	&	71.4	&	\textbf{75.85}	&	73.61	\\
    \textit{Neutral} &	87.62	&	87.55	&	\textbf{88.11}	&	87.83	\\
    
    \bottomrule
    \end{tabular}
    }
\end{table}

\section{Results}
\label{sec:results}
In this work, we used weighted f1-score as metric (higher the better) to measure emotion model classification performance.
We first show the effectiveness of \textit{CopyPaste} schemes on clean data and noise augmented data.
Then, we present results on artificially created noisy test data to compare \textit{CopyPaste} and noise augmentation.

\textit{General considerations:} Tables~\ref{tab:random_init_resnet_emotion_results} and~\ref{tab:pretrained_resnet_emotion_results} show the results of \textit{CopyPaste} schemes on randomly initialized ResNet model and speaker pre-trained ResNet model respectively.
Comparing both tables, we can observe that pre-training improves the model performance significantly on all datasets as is the case in~\cite{pappagari2020x}.
Models trained with noise augmented data perform better compared to models trained only on clean data corroborating with previous research~\cite{pappagari2020x,lakomkin2018robustness}.
Comparison of models trained with and without \textit{CopyPaste} schemes (4th-6th columns vs. 3rd column) reveals that our models perform better on all datasets with all schemes.
Though application of \textit{CopyPaste} schemes provide performance improvement in most cases, we do not observe a single best scheme across datasets and models except on Crema-D where \textit{N-CP} scheme consistently performs best.
We can observe that \textit{CopyPaste} schemes are effective on both clean data as well as noise augmented data.
We note that the improvements obtained with \textit{CopyPaste} schemes on the randomly initialized ResNet model are relatively higher compared to the improvements on the pre-trained ResNet model.

\textit{Per-class analysis:} As noted in the Section~\ref{subsec:emotion_recog}, there is a risk that the model can get biased to not predict neutral when \textit{N-CP} scheme is employed during model training.
Hence, we examined class-wise f1-scores of our models to identify the main source of improvements and observed that in most cases performance improved for all emotion classes.
As an example, we show in Table~\ref{tab:classwise_emotion_results} class-wise f1-scores of emotion classes on Crema-D dataset.
These scores are obtained with ResNet model pre-trained for speaker classification and trained on clean data.
We can observe improvements for all emotion classes with \textit{CopyPaste} schemes during training.
Among \textit{CopyPaste} schemes, \textit{N-CP} is performing best for all classes except for \textit{sad} emotion for which \textit{N+SE-CP} performs best.

\textit{Noise augmentation:} Comparing the augmentation techniques, \textit{CopyPaste} and noise augmentation, we can observe from Tables~\ref{tab:random_init_resnet_emotion_results} and~\ref{tab:pretrained_resnet_emotion_results} that \textit{CopyPaste} schemes performs better in most cases suggesting that concatenating utterances based on emotion helps the model generalize better compared to adding noise to the training data.
As noted in Section~\ref{subsubsec:noise_augmentation}, we compare noise augmentation and \textit{CopyPaste} in noisy test conditions too.
Tables~\ref{tab:emotion_results_SNR10} and~\ref{tab:emotion_results_SNR0} show the results on the noisy test data with SNR level of 10 dB and 0 dB respectively.
We used the model pre-trained with speaker classification for this experiment as it is performing the best on all the datasets.
As expected, SER performance degraded on the noisy test data suggesting that our models are sensitive to noisy test conditions.
Models trained with noise augmentation are more robust compared to models trained with only clean data which illustrates the benefits of augmenting training data with noise.
We can also observe that noise augmentation, in most cases, outperforms \textit{CopyPaste} in the noisy conditions.
However, our best models on all the datasets are when used both augmentations together which showcases the effectiveness of proposed  \textit{CopyPaste} schemes even in noisy test conditions.

\begin{table}\caption{SER results (weighted f1-scores) on noisy test data with $SNR=10dB$ with ResNet model pre-trained for speaker classification. \textit{Clean+Noise} and \textit{Clean} denote SER model training is on clean and augmented data, and clean data respectively; No \textit{CP} denotes model trained without \textit{CopyPaste}}
    \label{tab:emotion_results_SNR10}
    \centering
    \resizebox{\columnwidth}{!}{
    \begin{tabular}{l|l|l|l|l|l}
      \toprule
    Dataset    &Emotion data	&		No \textit{CP}	&	\textit{SE-CP} & 	\textit{N-CP}	&	\textit{N+SE-CP} 	\\
        	\midrule
    \multirow{2}{*}{MSP-Podcast} & Clean  	&55.25	&	\textbf{57.39}	&	55.54	&	56.61	\\
    & Clean+Noise 	& 57.09	&	\textbf{57.52}	&	56.63	&	\textbf{57.52}	\\
    \midrule
    \multirow{2}{*}{Crema-D}  & Clean 	& 72.76	&	73.47	&\textbf{77.06}	&	74.06	\\
    
     & Clean+Noise 	&	78.48	&	78.79	&	79.10	&	\textbf{79.30}	\\
    \midrule
    \multirow{2}{*}{IEMOCAP}  & Clean 	& 58.82	&	\textbf{59.30}	&	58.93	&	59.01	\\	
    & Clean+Noise 	&	61.47	&	61.80	&	62.03	&	\textbf{62.63}	\\	
    \bottomrule
    \end{tabular}
    }
\end{table}

\begin{table}\caption{SER results (weighted f1-scores) on noisy test data with $SNR=0dB$ with ResNet model pre-trained for speaker classification. \textit{Clean+Noise} and \textit{Clean} denote SER model training is on clean and augmented data, and clean data respectively; No \textit{CP} denotes model trained without \textit{CopyPaste}}
    \label{tab:emotion_results_SNR0}
    \centering
    \resizebox{\columnwidth}{!}{
    \begin{tabular}{l|l|l|l|l|l}
      \toprule
    Dataset    &Emotion data	&		No \textit{CP}	&	\textit{SE-CP} & 	\textit{N-CP}	&	\textit{N+SE-CP} 	\\
        	\midrule
    \multirow{2}{*}{MSP-Podcast} &  Clean  	& 52.65	&	\textbf{55.15}	&	52.68	&	53.91	\\
    & Clean+Noise 	& 55.88	&	56.40	&	55.28	&	\textbf{56.44}	\\
    \midrule
    \multirow{2}{*}{Crema-D}  & Clean 	& 64.95	&	66.21	&	\textbf{71.40}	&	65.53	\\
    
     & Clean+Noise 	&	76.44	&	76.38	&	\textbf{76.83}	&	76.60	\\
    \midrule
    \multirow{2}{*}{IEMOCAP}  & Clean 	& \textbf{52.05}	&	51.73	&	51.58	&	51.62	\\
    & Clean+Noise 	&	58.55	&	58.52	&	59.32	&	\textbf{59.69}	\\
    \bottomrule
    \end{tabular}
    }
\end{table}

\section{Conclusions}
\label{sec:conclusion}
In this work, we proposed \textit{CopyPaste}, a perceptually motivated augmentation technique, for speech emotion recognition (SER) task.
We postulated that for emotions other than neutral, a speaker is perceived to be expressing an emotion $E$ even if it is expressed within a short segment in a longer utterance.
We presented three \textit{CopyPaste} schemes for model training based on the concatenation of utterances w.r.t. emotion labels.
We have evaluated \textit{CopyPaste} augmentation on 3 contrasting emotion datasets using 2 neural net (ResNet) models: one trained independently and another using transfer learning from a model pre-trained to classify speakers.
We have shown that all schemes of \textit{CopyPaste} improve the SER on all the datasets and perform better than standard speech augmentation technique of adding noise to the signal.
Our results suggested that \textit{CopyPaste} is effective even in noisy test conditions.
Our best performing models use both \textit{CopyPaste} and noise augmentation for training.
We note that even though \textit{CopyPaste} schemes are presented for SER task in this paper, the underlying idea can be applied to other utterance based classification tasks such as language identification and age prediction.

\bibliographystyle{IEEEbib}
\bibliography{main}

\begin{thebibliography}{10}

\bibitem{koolagudi2012emotion}
Shashidhar~G Koolagudi and K~Sreenivasa Rao,
\newblock ``Emotion recognition from speech: a review,''
\newblock {\em International journal of speech technology}, vol. 15, no. 2, pp.
  99--117, 2012.

\bibitem{arevian2020clinical}
Armen~C Arevian, Daniel Bone, Nikolaos Malandrakis, Victor~R Martinez,
  Kenneth~B Wells, David~J Miklowitz, and Shrikanth Narayanan,
\newblock ``Clinical state tracking in serious mental illness through
  computational analysis of speech,''
\newblock {\em PLoS one}, vol. 15, no. 1, pp. e0225695, 2020.

\bibitem{lopez2015feature}
Karmele Lopez-de Ipi{\~n}a, JB~Alonso-Hern{\'a}ndez, Jordi Sol{\'e}-Casals,
  Carlos~Manuel Travieso-Gonz{\'a}lez, Aitzol Ezeiza, Marcos Faundez-Zanuy,
  Pilar~M Calvo, and Blanca Beitia,
\newblock ``Feature selection for automatic analysis of emotional response
  based on nonlinear speech modeling suitable for diagnosis of alzheimer's
  disease,''
\newblock {\em Neurocomputing}, vol. 150, pp. 392--401, 2015.

\bibitem{schmidt2017survey}
Anna Schmidt and Michael Wiegand,
\newblock ``A survey on hate speech detection using natural language
  processing,''
\newblock in {\em Proceedings of the Fifth International workshop on natural
  language processing for social media}, 2017, pp. 1--10.

\bibitem{huang2014speech}
Zhengwei Huang, Ming Dong, Qirong Mao, and Yongzhao Zhan,
\newblock ``Speech emotion recognition using cnn,''
\newblock in {\em Proceedings of the 22nd ACM international conference on
  Multimedia}. ACM, 2014, pp. 801--804.

\bibitem{lim2016speech}
Wootaek Lim, Daeyoung Jang, and Taejin Lee,
\newblock ``Speech emotion recognition using convolutional and recurrent neural
  networks,''
\newblock in {\em 2016 Asia-Pacific Signal and Information Processing
  Association Annual Summit and Conference (APSIPA)}. IEEE, 2016, pp. 1--4.

\bibitem{cho2018deep}
Jaejin Cho, Raghavendra Pappagari, Purva Kulkarni, Jes{\'u}s Villalba, Yishay
  Carmiel, and Najim Dehak,
\newblock ``Deep neural networks for emotion recognition combining audio and
  transcripts.,''
\newblock in {\em Interspeech}, 2018, pp. 247--251.

\bibitem{zhao2019speech}
Jianfeng Zhao, Xia Mao, and Lijiang Chen,
\newblock ``Speech emotion recognition using deep 1d \& 2d cnn lstm networks,''
\newblock {\em Biomedical Signal Processing and Control}, vol. 47, pp.
  312--323, 2019.

\bibitem{eyben2013recent}
Florian Eyben, Felix Weninger, Florian Gross, and Bj{\"o}rn Schuller,
\newblock ``Recent developments in opensmile, the munich open-source multimedia
  feature extractor,''
\newblock in {\em Proceedings of the 21st ACM international conference on
  Multimedia}. ACM, 2013, pp. 835--838.

\bibitem{tzirakis2017end}
Panagiotis Tzirakis, George Trigeorgis, Mihalis~A Nicolaou, Bj{\"o}rn~W
  Schuller, and Stefanos Zafeiriou,
\newblock ``End-to-end multimodal emotion recognition using deep neural
  networks,''
\newblock {\em IEEE Journal of Selected Topics in Signal Processing}, vol. 11,
  no. 8, pp. 1301--1309, 2017.

\bibitem{sarma2018emotion}
Mousmita Sarma, Pegah Ghahremani, Daniel Povey, Nagendra~Kumar Goel,
  Kandarpa~Kumar Sarma, and Najim Dehak,
\newblock ``Emotion identification from raw speech signals using dnns.,''
\newblock in {\em Interspeech}, 2018, pp. 3097--3101.

\bibitem{trigeorgis2016adieu}
George Trigeorgis, Fabien Ringeval, Raymond Brueckner, Erik Marchi, Mihalis~A
  Nicolaou, Bj{\"o}rn Schuller, and Stefanos Zafeiriou,
\newblock ``Adieu features? end-to-end speech emotion recognition using a deep
  convolutional recurrent network,''
\newblock in {\em 2016 IEEE international conference on acoustics, speech and
  signal processing (ICASSP)}. IEEE, 2016, pp. 5200--5204.

\bibitem{lakomkin2018reusing}
Egor Lakomkin, Cornelius Weber, Sven Magg, and Stefan Wermter,
\newblock ``Reusing neural speech representations for auditory emotion
  recognition,''
\newblock {\em arXiv preprint arXiv:1803.11508}, 2018.

\bibitem{pappagari2020x}
Raghavendra Pappagari, Tianzi Wang, Jesus Villalba, Nanxin Chen, and Najim
  Dehak,
\newblock ``x-vectors meet emotions: A study on dependencies between emotion
  and speaker recognition,''
\newblock in {\em ICASSP 2020-2020 IEEE International Conference on Acoustics,
  Speech and Signal Processing (ICASSP)}. IEEE, 2020, pp. 7169--7173.

\bibitem{hestness2017deep}
Joel Hestness, Sharan Narang, Newsha Ardalani, Gregory Diamos, Heewoo Jun,
  Hassan Kianinejad, Md~Patwary, Mostofa Ali, Yang Yang, and Yanqi Zhou,
\newblock ``Deep learning scaling is predictable, empirically,''
\newblock {\em arXiv preprint arXiv:1712.00409}, 2017.

\bibitem{lakomkin2018robustness}
Egor Lakomkin, Mohammad~Ali Zamani, Cornelius Weber, Sven Magg, and Stefan
  Wermter,
\newblock ``On the robustness of speech emotion recognition for human-robot
  interaction with deep neural networks,''
\newblock in {\em 2018 IEEE/RSJ International Conference on Intelligent Robots
  and Systems (IROS)}. IEEE, 2018, pp. 854--860.

\bibitem{etienne2018cnn+}
Caroline Etienne, Guillaume Fidanza, Andrei Petrovskii, Laurence Devillers, and
  Benoit Schmauch,
\newblock ``Cnn+ lstm architecture for speech emotion recognition with data
  augmentation,''
\newblock {\em arXiv preprint arXiv:1802.05630}, 2018.

\bibitem{bao2019cyclegan}
Fang Bao, Michael Neumann, and Ngoc~Thang Vu,
\newblock ``Cyclegan-based emotion style transfer as data augmentation for
  speech emotion recognition.,''
\newblock in {\em INTERSPEECH}, 2019, pp. 2828--2832.

\bibitem{rizos2020stargan}
Georgios Rizos, Alice Baird, Max Elliott, and Bj{\"o}rn Schuller,
\newblock ``Stargan for emotional speech conversion: Validated by data
  augmentation of end-to-end emotion recognition,''
\newblock in {\em ICASSP 2020-2020 IEEE International Conference on Acoustics,
  Speech and Signal Processing (ICASSP)}. IEEE, 2020, pp. 3502--3506.

\bibitem{toth2008speech}
Szabolcs~Levente T{\'o}th, David Sztah{\'o}, and Kl{\'a}ra Vicsi,
\newblock ``Speech emotion perception by human and machine,''
\newblock in {\em Verbal and Nonverbal Features of Human-Human and
  Human-Machine Interaction}, pp. 213--224. Springer, 2008.

\bibitem{villalba2019state}
Jes{\'u}s Villalba, Nanxin Chen, David Snyder, Daniel Garcia-Romero, Alan
  McCree, Gregory Sell, Jonas Borgstrom, Fred Richardson, Suwon Shon,
  Fran{\c{c}}ois Grondin, et~al.,
\newblock ``State-of-the-art speaker recognition for telephone and video
  speech: The jhu-mit submission for nist sre18,''
\newblock {\em Proc. Interspeech 2019}, pp. 1488--1492, 2019.

\bibitem{he2016deep}
Kaiming He, Xiangyu Zhang, Shaoqing Ren, and Jian Sun,
\newblock ``Deep residual learning for image recognition,''
\newblock in {\em Proceedings of the IEEE conference on computer vision and
  pattern recognition}, 2016, pp. 770--778.

\bibitem{snyder2015musan}
David Snyder, Guoguo Chen, and Daniel Povey,
\newblock ``Musan: A music, speech, and noise corpus,''
\newblock {\em arXiv preprint arXiv:1510.08484}, 2015.

\bibitem{hsiao2015robust}
Roger Hsiao, Jeff Ma, William Hartmann, Martin Karafi{\'a}t, Franti{\v{s}}ek
  Gr{\'e}zl, Luk{\'a}{\v{s}} Burget, Igor Sz{\"o}ke, Jan~Honza
  {\v{C}}ernock{\`y}, Shinji Watanabe, Zhuo Chen, et~al.,
\newblock ``Robust speech recognition in unknown reverberant and noisy
  conditions,''
\newblock in {\em 2015 IEEE Workshop on Automatic Speech Recognition and
  Understanding (ASRU)}. IEEE, 2015, pp. 533--538.

\bibitem{lotfian2017building}
Reza Lotfian and Carlos Busso,
\newblock ``Building naturalistic emotionally balanced speech corpus by
  retrieving emotional speech from existing podcast recordings,''
\newblock {\em IEEE Transactions on Affective Computing}, 2017.

\bibitem{busso2008iemocap}
Carlos Busso, Murtaza Bulut, Chi-Chun Lee, Abe Kazemzadeh, Emily Mower, Samuel
  Kim, Jeannette~N Chang, Sungbok Lee, and Shrikanth~S Narayanan,
\newblock ``Iemocap: Interactive emotional dyadic motion capture database,''
\newblock {\em Language resources and evaluation}, vol. 42, no. 4, pp. 335,
  2008.

\end{thebibliography}

\end{document}